\documentclass[10pt,aps,prc,twocolumn,amssymb,showpacs,floatfix]{revtex4-1}

\usepackage{graphicx}
\usepackage{amsmath}
\usepackage{float} 
\usepackage{breqn} 
\usepackage[british]{babel}
\usepackage{algpseudocode} 
\usepackage{multirow} 
\usepackage[breaklinks,colorlinks]{hyperref} 

\newcommand{\bref}[1]{(\ref{#1})}

\newcommand{\deriv}[2]{\frac{d #1}{d #2}}

\newcommand{\pderiv}[2]{\frac{\partial #1}{\partial #2}}
\newcommand{\pderivtwo}[2]{\frac{\partial^2 #1}{\partial #2^2}}

\newcommand{\brac}[1]{\left( #1\right)}
\newcommand{\sbrac}[1]{\left[ #1\right]}
\newcommand{\braket}[2]{\langle #1\mid #2\rangle}
\newcommand{\braHket}[3]{\langle #1\mid#2\mid #3\rangle}

\newcommand{\half}[0]{\frac{1}{2}}
\newcommand{\TDSEco}[0]{-\frac{\hbar^2}{2m}}

\newcommand{\TISEarg}[0]{\vec{z_1},...,\vec{z_N}}

\newcommand{\chem}[3]{
\ensuremath{
\phantom{\ensuremath{^{#1}_{#2}}}
\llap{\ensuremath{^{#1}}}
\llap{\ensuremath{_{\rule{0pt}{.75em}#2}}}
\mbox{#3}
}
}

\hypersetup{citecolor=blue}
\begin{document}

\floatstyle{ruled}
\newfloat{algorithm}{h}{algs}
\floatname{algorithm}{Algorithm }

\title{Extension of continuum time-dependent Hartree-Fock method to proton states}
\author{C. I. Pardi}
\email{c.pardi@surrey.ac.uk}
\author{P. D. Stevenson}
\email{p.stevenson@surrey.ac.uk}
\affiliation{Department of Physics$,$ University of Surrey$,$ Guildford$,$ Surrey$,$ GU2 7XH$,$ United Kingdom}
\author{K. Xu}
\email{kuan.xu@maths.ox.ac.uk}
\affiliation{Mathematical Institute$,$ University of Oxford$,$ Oxford$,$ Oxfordshire$,$ OX1 3LB$,$ United Kingdom}
\date{\today}

\begin{abstract}
This paper deals with the solution of the spherically symmetric time-dependent Hartree-Fock approximation applied to nuclear giant monopole resonances in the small amplitude regime. The problem is spatially unbounded as the resonance state is in the continuum. The practical requirement to perform the calculation in a finite-sized spatial region yields an artificial boundary, which is not present physically. The question of how to ensure the boundary does not interfere with the internal solution, while keeping the overall calculation time low is studied. Here we propose an absorbing boundary condition scheme to handle the conflict. The derivation, via a Laplace transform method, and implementation is described. An inverse Laplace transform required by the absorbing boundaries is calculated using a method of non-linear least squares. The accuracy and efficiency of the scheme is tested and results presented to support the case that they are a effective way of handling the artificial boundary.
\end{abstract}


\maketitle
 
\section{\label{sec:Introduction}Introduction}

As a quantum system the behaviour of a nucleus over a period of time obeys the $N$-particle time-dependent Schroedinger equation (TDSE). Solving the full many-body TDSE analytically or even numerically is generally not tractable. However, approximate solutions can be gained by solving the time dependent Hartree Fock (TDHF) equations \cite{ring2005nuclear,nesbetvariational}. The simplification still does not allow analytic solutions, but numerical techniques can be applied and the computational cost kept manageable.

The TDHF equations are a coupled set of initial-boundary-value problems for which it is common to apply finite differencing methods in both spatial and temporal coordinates.  In this scheme the equations can be solved by e.g. a series of matrix inversions. One difficulty with these types of computational solution is the limitation of calculating wave functions in a finite spatial region, which introduces an artificial boundary into calculations. Appropriate conditions for the boundary have to be chosen. In cases where the system can be fully contained in a region for all time, the values at the boundary can simply and correctly be set to zero. However, in many situations particles are emitted from a system into the continuum.  This is common in the case of giant resonances, as most are above the particle decay threshold \cite{harakeh2001giant}. These particles move off into the continuum physically, yet computationally they will reach the artificial boundary \cite{springerlink:10.1140/epja/i2007-10366-9,PhysRevC.69.054322}.   

The most crude, and simple, way of tackling this problem is to use reflecting boundaries, which rebound any matter that comes into contact with them. A fully accurate solution with these boundary conditions can occur in the case that the reflecting boundary is sufficiently distant from the original nucleus that emitted flux does not reach the boundary within the simulation time.  Although such reflecting boundaries are easy to implement, in some cases the large spatial domain required results in inefficient calculations \cite{giantRes,springerlink:10.1140/epjad/i2005-06-052-x}.

More efficient solutions have been sought in the form of absorbing potentials. These attempt to use reflecting boundaries in a sensibly sized region, and then use a complex potential to remove matter that approaches the boundary. This can prevent reflections taking place and work well in some situations \cite{springerlink:10.1140/epjad/i2005-06-052-x}. However, in general a given absorbing potential will not work with perfect efficiency at all frequencies, and these potentials may still require considerable spatial extent to work very well \cite{giantRes,PhysRevC.71.024301}.

Here we present a method of implementing absorbing boundary conditions (ABCs) \cite{antoine2009review}. These rely on choosing the artificial boundary such that the potential outside of it has a simple form. The resulting equations in the exterior can be manipulated into a boundary condition for the interior problem which can be applied closer to the initial bulk of matter. The propagation of waves in the exterior region then does not have to be dealt with explicitly. In solving the TDHF equations, a simplified Skyrme interaction is used in the implementation which reproduces the magic numbers needed for $\chem{4}{2}{He}$, $\chem{16}{8}{O}$, and $\chem{40}{20}{Ca}$ to be seen without the complexity of the full interaction \cite{PhysRevC.60.044302}, as a reasonable proof-of-concept. Spherical symmetry is also assumed inside and outside of the artificial boundary. The calculations involve various forms of differential equation, each of which requiring their own absorbing boundary conditions. Here a continuous absorbing boundary condition is implemented \cite{antoine2009review} which improves on previous work \cite{PhysRevC.87.014330} by accounting for the long-ranged Coulomb potential. The previous work is modified by approximation of the required inverse Laplace transform via the use of a non-linear least squares method \cite{XuJiangBootstrap2013}.

The structure of this paper is as follows. Section \ref{sec:Giant Monopole Resonances} gives a brief summary of the types of giant resonance and their properties. The theory and discretization for the time-dependent Hartree-Fock approach is described in section \ref{sec:Time-Dependent Hartree-Fock}. Section \ref{sec:Boundary Conditions} describes the absorbing boundary conditions, the non-linear least squares method and their application to TDHF. Testing of the ABCs implementation is given in section \ref{sec:Testing} and results of the TDHF with ABCs calculations are given in section \ref{sec:Results}

\section{\label{sec:Giant Monopole Resonances}Giant Monopole Resonances}

Giant monopole resonances (GMRs) are collective excitations of the nucleus, meaning most if not all particles are involved in the excitation \cite{harakeh2001giant}. They are well studied experimentally, being first observed in 1977 \cite{PhysRevLett.38.676} and their study has continued to the present day \cite{Patel2012447,PhysRevC.23.1997,PhysRevC.22.1832}. Excitation of the monopole resonance is commonly performed with $\alpha$-scattering \cite{Patel2012447}. The requirement for angular momentum conservation excludes the possibility of excitation by a photon, as is performed for the dipole resonant mode \cite{RevModPhys.47.713}.  Aside from shedding light on the structure of individual nuclei, further interest in GMRs is provided by their relation to the incompressibility of nuclear matter and the consequent light they shed on the equation of state with consequent importance in understanding neutron stars, supernovae explosions and heavy-ion collision \cite{PhysRevLett.82.691}.

Our main interest in this phenomenon, however, is owing to the simplified analysis they allow for. Specifically they are a purely radial excitation and hence by considering only the subset of doubly magic nuclei we are able take advantage of spherical symmetry in the calculations. As is common when developing new methods a simplified Skyrme potential, containing just the $t_0$ and $t_3$ terms, is used \cite{PhysRevC.60.044302,Koonin1976227,Stringari1979,Stevenson2010,Almehed2005,0954-3899-31-10-079}. As was commented on previously \cite{PhysRevC.87.014330} this cannot be expected to gives a detailed comparison with experiment, but is used to demonstrate the features of the new method.

The key quantity for comparison to experiment is the strength function, which can be related to the experimental cross section. This quantity has been noted to be particularly sensitive to the boundary conditions applied to the TDHF equations \cite{giantRes}. Therefore, we shall measure success as the accurate reproduction of this quantity, free of artefacts that may arise from the boundary conditions.

\section{\label{sec:Time-Dependent Hartree-Fock}Time-Dependent Hartree-Fock (TDHF)}

Originating with a formulation by Dirac \cite{Dirac1930}, the time-dependent Hartree-Fock method became practical for realistic calculations in nuclei only with the advent of sufficiently advanced computational facilities \cite{Bonche1976,Cusson1976,Davies1985}.  It has been widely applied to heavy-ion collisions and giant resonances, as well as selected other problems.  A recent review \cite{Simenel2012} covers many such applications.

\subsection{\label{subsec:Theory}Theory}

The TDHF method relies on the time dependent variational principle in which the action, defined as
\begin{eqnarray}
S[\Psi(t)] = \int_{t_0}^{t_1} \braHket{\Psi(t)}{i\hbar\pderiv{}{t} - \hat{H}}{\Psi(t)\,} \, dt, \label{eqn:4-ActionBraKet}
\end{eqnarray}
is minimized. If one considers a trial wavefunction $\mid\Psi(t)\rangle$ belonging to a general Hilbert space it can be shown that the Schroedinger equation is retrieved upon minimising the above. The TDHF method considers a trial wavefunction in a restricted space of antisymmetric Slater determinants \cite{gross1991many}, given in the spatial-spin-isospin basis as
\begin{dmath}[compact]
\Psi^{(A)}(\vec{z_1},\hdots,\vec{z_N},t) 
=
\frac{1}{\sqrt{N!}}
\begin{vmatrix}
	\phi_{1}(\vec{z_1},t) & \hdots & \phi_{1}(\vec{z_N},t)\\
	\vdots & \ddots & \vdots\\
	\phi_{N}(\vec{z_1},t) & \hdots & \phi_{N}(\vec{z_N},t)\\
\end{vmatrix}.
\end{dmath}
The coordinate $\vec{z}_i=\brac{\vec{r}_i,\sigma_i,\tau_i}$ describes spatial, spin and isospin degrees of freedom. The wanted result from minimising in this space of restricted wavefunctions is to produce a numerically tractable problem. The Hamiltonian, $\hat{H}$, for nuclear calculations is accepted to contain a kinetic operator and two and three body operators that describe the potential \cite{ring2005nuclear,greiner1996nuclear,RevModPhys.54.913}. In the spatial-spin-isospin basis this takes the form
\begin{dmath}[compact]
\hat{H}(\TISEarg) = \TDSEco\sum_{i=1}^{N}\nabla^2_i(\vec{r_i}) 
+ \sum_{i=1}^{N}\sum_{j=1}^{i-1} \hat{v}^{(2)}_{ij}(\vec{z_i},\vec{z_j})
+ \sum_{i=1}^{N}\sum_{j=1}^{i-1}\sum_{k=1}^{j-1} \hat{v}^{(3)}_{ijk}(\vec{z_i},\vec{z_j},\vec{z_k}).
\label{eqn:Many-Body Schroedinger equation full}
\end{dmath}
In this work we shall use the simplified $t_0-t_3$ Skyrme interaction for the nuclear components of the potential and the electrostatic interaction for the Coulomb component. This yields the two body potential as \cite{Skyrme1958,doi:10.1080/14786435608238186}
\begin{eqnarray}
v_{ij}^{(2)}(\vec{r},\vec{r'}) 
=t_0\delta\brac{\vec{r} - \vec{r'}} + \frac{\eta}{|\vec{r}-\vec{r'}|}P_{i,j} \label{eqn:3-v2potential}
\end{eqnarray}
where $\eta = \frac{e^2}{4\pi\epsilon_0}\approx 1.44$ e$^2$ MeV$^{-1}$ fm$^{-1}$ and $P_{ij}$ is zero if $i$ and or $j$ is a neutron and one if $i$ and $j$ are protons. The three body potential is given as \cite{Skyrme1958,doi:10.1080/14786435608238186}
\begin{eqnarray}
v^{(3)}_{ijk}(\vec{r},\vec{r'},\vec{r''}) = t_3\delta(\vec{r} - \vec{r'})\delta(\vec{r'} - \vec{r''}). \label{eqn:3-threeBodyPot}
\end{eqnarray}
The values $t_0=-1090.0$ MeV fm$^3$ and $t_3=17288.0$ MeV fm$^6$ are used \cite{PhysRevC.87.014330}. Performing the minimization of the action with the Hamiltonian as described above can be shown to produce the following set of equations for the reduced radial single particle wavefunctions
\begin{dmath}[compact]
i\hbar\pderiv{Q_{n,l}(r,t)}{t}
=
{\hat{H}_{HF}} {Q_{n,l}}(r,t),
\label{eqn:3-TDHF SS DE}
\end{dmath}
where the Hartree-Fock Hamiltonian is given as
\begin{dmath}[compact]
\hat{H}_{HF} = \sbrac{\TDSEco\pderiv{}{r^2}
+ V(r,t,\rho_{\text{n}},\rho_{\text{p}})
+ \frac{\hbar^2}{2m}\frac{l(l+1)}{r^2} }.
\label{eqn:Hartree-Fock Hamiltonian}
\end{dmath}
The above equations are subject to the boundary conditions;
\begin{eqnarray}
&&Q_{n,l}(0,t) = 0, \label{eqn: TDHF BC 1} \\
&&Q_{n,l}(r,t) \to 0,\mbox{\quad as } r\to\infty, \label{eqn: TDHF BC 2}
\end{eqnarray}
and an initial condition, described later. The spatial part of the three-dimensional single particle wavefunctions can be retrieved from the above from
\begin{eqnarray}
\phi(\vec{r},t)=\frac{Q_{n,l}(r,t)}{r}Y_l^m(\theta,\varphi),
\label{eqn:TD Single Particle WF Seperate}
\end{eqnarray}
where $Y_l^m(\theta,\varphi)$ is a spherical harmonic and $l$ and $m$ are the orbital and magnetic quantum numbers respectively. When calculating a neutron single particle wavefunction the potential $V(r,t,\rho_{\text{n}},\rho_{\text{p}})$ is equal to
\begin{dmath}
V_{\text{n}}(r,t,\rho_{\text{n}},\rho_{\text{p}}) 
= 
t_0\brac{\rho_{\text{p}}+\half\rho_{\text{n}}}
+
\frac{t_3}{4}\rho_{\text{p}}\brac{\rho_{\text{p}} + 2\rho_{\text{n}}}
\label{eqn:TDHF Neutron Potential}
\end{dmath}
and when calculating a proton
\begin{dmath}
V_{\text{p}}(r,t,\rho_{\text{n}},\rho_{\text{p}}) 
=
t_0\brac{\rho_{\text{n}}+\half\rho_{\text{p}}} 
+ \frac{t_3}{4}\rho_{\text{n}}\brac{\rho_{\text{n}} + 2\rho_{\text{p}}}
+ V_c(r,t).
\label{eqn:TDHF Proton Potential}
\end{dmath}
The densities are given by
\begin{eqnarray}
\rho(r,t) 
= \frac{1}{4\pi r^2}\sum_{(n,l)\in S} (2l+1)\left|Q_{n,l}(r,t) \right|^2,
\label{eqn:TDHF Density}
\end{eqnarray}
where the sets $S_{\text{n}}$ and $S_{\text{p}}$ replace $S$, in the above, for the neutron and proton densities, $\rho_{\text{n}}$ and $\rho_{\text{p}}$, respectively. The sets $S_{\text{p}}$ and $S_{\text{n}}$ contains the values of $(n,l)$ for the protons and neutrons within the system. The particular $(n,l)$ values we take for each nuclei are shown in table \ref{table:3-NucleiStates}.

\begin{table}[h]
\centering
\begin{tabular*}{0.45\textwidth}{@{\extracolsep{\fill} }   l c c }
\hline\hline
	Nucleus &  $(n,l)\in S_{\text{n}}$ or $S_{\text{p}}$ & $2l+1$ \\
\hline
	Helium-4 & (0,0),(0,0) & 1\\
\hline
	Oxygen-16 & (0,0),(0,0) & 1 \\
					  & (0,1),(0,1) & 3\\
\hline
Calcium-40   & (0,0),(0,0) & 1\\
	 				  & (1,0),(1,0) & 1\\
	 				  & (0,1),(0,1) & 3\\
					  & (0,2),(0,2) & 5\\
\hline\hline
\end{tabular*}
\caption{Table showing the explicit elements of the set $S_{\text{n}}$ and $S_{\text{p}}$ for the nuclei considered here.}
\label{table:3-NucleiStates}
\end{table}

The Coulomb potential, $V_c(r,t)$ can be found by solving the following differential equation:
\begin{eqnarray}
\pderivtwo{W_c(r,t)}{r}=-4\pi\eta r\rho_{\text{p}}(r,t),
\label{eqn:3-SS TD Coulomb DE}
\end{eqnarray}
for $W_c$, subject to the boundary conditions;
\begin{eqnarray}
W_c(0,t) = 0, \label{eqn: TD Coulomb BC 1}\\
\left.\pderiv{W_c(r,t)}{r}\right|_{r=R_{\text{Coul}}} = 0. \label{eqn: TD Coulomb BC 2}
\end{eqnarray}
where $\rho_{\text{p}}=0$ for  $r \ge R_{\text{Coul}}$. The Coulomb potential can then be calculated from $W_c(r,t)$ via
\begin{eqnarray}
V_c(r,t) = \frac{W_c(r,t)}{r}. \label{eqn:V_c from W_c}
\end{eqnarray}
It is noted that the minimisation also produces an exchange term for the Coulomb potential which is excluded in this analysis.

\subsubsection{\label{subsec:The Initial Condition}The Initial Condition}

The initial condition in these calculation is defined to be the result from applying a boost operator on the ground state
\begin{eqnarray}
\Psi(\vec{z_1},\hdots,\vec{z_N},t=0)  &=&
e^{i\epsilon r^2} \Psi_0(\vec{z_1},\hdots,\vec{z_N}) \label{eqn:2-resonanceOperator}
\end{eqnarray}
The ground state, $\Psi_0(\vec{z_1},\hdots,\vec{z_N})$, is found using the time-independent Hartree-Fock method in which the energy is minimized in a space of Slater determinants to produce
\begin{eqnarray}
{\hat{H}_{HF}}{Q_{n,l}}(r) = {E_{n,l}}{Q_{n,l}}(r),
\label{eqn:Stationary Hartree Fock equation}
\end{eqnarray}
which relies on the the time-independent equivalents of equations \bref{eqn:Hartree-Fock Hamiltonian} to \bref{eqn:V_c from W_c}.

\subsection{\label{subsec:Numerical Procedure}Numerical Procedure}

Equations \bref{eqn:3-TDHF SS DE}, \bref{eqn:3-SS TD Coulomb DE} and \bref{eqn:Stationary Hartree Fock equation} are all solved numerically by finite difference methods. So the following discrete spatial variable is defined:
\begin{eqnarray}
r_m \equiv m\Delta r \label{eqn:discete grid}, \\
\Delta r=\frac{R}{M} \nonumber,
\end{eqnarray}
where $m=1,2,\hdots,M$. Time is chosen to be discretised by the equidistant set of points,
\begin{eqnarray}
t_n = n\Delta t,\\
\Delta t=\frac{T}{N} \nonumber,
\end{eqnarray}
where $n=0,1,\hdots,N$. 

We use the methods described previously \cite{PhysRevC.87.014330} to calculate the ground state and time-dependent wavefunction on the spatial and temporal grid. Linear equations are produced for the stationary case through use of a self-consistent scheme, which can be expressed as a series of matrix eigen-value problems by using central differences. An intermediate step via the evolution operator is used to produce linear equations in the time-dependent case, which are then discretised in time using the Crank-Nicholson scheme \cite{crankNic} and space using central differences, producing a series of matrix inversions. In practise the Lapack subroutines \cite{lapack1999} are used to solve the matrix equations.

\section{\label{sec:Boundary Conditions}Boundary Conditions}

In this section we discuss the method of treating the TDHF equations in the continuum. The first part of the section discusses the derivation of an absorbing boundary condition, applicable to nuclear calculations. This will be seen to require the inverse Laplace transform of a kernel. A non-linear least squares approach \cite{Jiang2001,XuJiangBootstrap2013} is then described to provide an accurate approximate of the kernel by a sum of poles, whose inversion can be found in tables \cite{AbraMathFunc,intTransErdelyi}. Finally, discretization of the absorbing boundary condition for use with the Crank-Nicholson scheme is described.

\subsection{\label{sec:The Problem in the Exterior}The Problem in the Exterior}

Application of absorbing boundary conditions require us to split the domain into two regions; an interior, and an exterior \cite{antoine2009review,Mayfield1989}. In the nuclear problem we have two types of potential; short ranged Skymre potentials, and the long ranged Coulomb and centrifugal potentials. We choose to split our domain so that the potential in the exterior contains just the long-ranged components.

In order to ensure the Skyrme potentials are only present in the interior, we assume that the density is zero in the exterior. So in our analysis we make the assumption
\begin{eqnarray}
\rho_{\text{p}} = \rho_{\text{n}} = 0,
\label{eqn:Density 0 Assumption}
\end{eqnarray}
for $r\ge R$. Of course this will only be approximately true, as some of the density will move into the exterior during the calculation. However, as was shown previously in the linear regime the effect is not detrimental to the results \cite{PhysRevC.87.014330}. Using the above we may write the potentials \bref{eqn:TDHF Neutron Potential} and \bref{eqn:TDHF Proton Potential} in the exterior as
\begin{eqnarray}
&&V_{\text{n}}(r,\rho_{\text{n}},\rho_{\text{p}},t) \label{eqn:exterior Neutron}
= 
0, \\
&&V_{\text{p}}(r,\rho_{\text{n}},\rho_{\text{p}},t) \label{eqn:exterior Proton}
= 
V_c(r,t).
\end{eqnarray}
Assumption \bref{eqn:Density 0 Assumption} and Gauss' Law also allows for the Coulomb potential to be simplified to
\begin{eqnarray}
V_c(r) =\frac{\eta N_{\text{p}}}{r}
\label{eqn: exterior Coulomb}.
\end{eqnarray}
Assuming elementary charge units, $N_{\text{p}}$ is the number of protons. Using equations \bref{eqn:exterior Neutron}, \bref{eqn:exterior Proton} and \bref{eqn: exterior Coulomb} with equation \bref{eqn:3-TDHF SS DE} we can write the following form of equation that all general single particle wavefunctions obey in the exterior:
\begin{dmath}[compact]
i\hbar\pderiv{Q(r,t)}{t} 
=
-\frac{\hbar^2}{2m}\pderivtwo{Q(r,t)}{r}  + \brac{\frac{\sigma}{r} + \frac{\hbar^2}{m}\frac{l(l+1)}{2r^2}}Q(r,t),
\label{eqn: SP exterior equation}
\end{dmath}
which is subject to the boundary condition
\begin{eqnarray}
\lim_{r\to\infty}Q(r,t) = 0.
\label{eqn: BC SP exterior}
\end{eqnarray}
Within equation \bref{eqn: SP exterior equation} we use
\begin{eqnarray}
\sigma
=
\begin{cases}
\eta N_{\text{p}}, & \text{for protons} \\
0, & \text{for neutrons}
\end{cases}.
\end{eqnarray}
The $n,l$ values are kept implicit in the notation for $Q$ as we are considering the form of a general single particle wavefunction. We continue by deriving some absorbing boundary conditions for Schroedinger-like equations that have the above form in an exterior region.

\subsection{\label{subsec:Absorbing Boundary Conditions}Absorbing Boundary Conditions}

Equation \bref{eqn: SP exterior equation} may be written more compactly by letting $t\to \frac{m}{\hbar}t$ and $\sigma\to \frac{m}{\hbar^2}\sigma$, producing
\begin{dmath}[compact]
i\pderiv{Q(r,t)}{t} = -\half\pderivtwo{Q(r,t)}{r} + \brac{\frac{\sigma}{r} + \frac{l(l+1)}{2r^2}}Q(r,t).
\label{eqn: SP exterior Dimen}
\end{dmath}
We now recall the definition of the Laplace transform,
\begin{eqnarray}
\hat{f}(s) = \int_0^\infty f(t) e^{-st} \, dt
\end{eqnarray}
and its inverse, the Bromwich integral \cite{transMethDuffy},
\begin{eqnarray}
f(t) = \frac{1}{2\pi i}\int_{c-i\infty}^{c+i\infty} \hat{f}(s)e^{st} \, ds.
\label{eqn:inversionDef}
\end{eqnarray}
$c$ is chosen such that the poles of $\hat{f}(s)$ are to the left of the contour. The hat notation is now used to imply the Laplace transform of a function. We proceed by multiplying equation \bref{eqn: SP exterior Dimen} by $e^{-st}$ and integrating in time from 0 to $\infty$, to get the differential equation for the Laplace transform of $Q(r,t)$
\begin{eqnarray}
\half\pderivtwo{\hat{Q}(r,s)}{r} + \brac{is -\frac{\sigma}{r} -\frac{l(l+1)}{2r^2} }\hat{Q}(r,s) = 0.
\label{eqn:LaplaceODE}
\end{eqnarray}
The above is simplified by assuming the initial condition is zero in the exterior region. This isn't restrictive for our needs because the nuclear wavefunction is localised around the origin. Letting $z=br\sqrt{s}$, where $b=-2i\sqrt{2i}$ and choosing the square root to be on the branch having positive real part, produces
\begin{dmath}[compact]
\pderivtwo{\hat{Q}(r,s)}{z} + \brac{-\frac14 +\frac{\kappa(s)}{z} -\frac{\frac14 - \mu^2}{z^2} }\hat{Q}(r,s) = 0
\label{eqn:Whittaker Equation}
\end{dmath}
where
\begin{eqnarray}
\kappa(s) = -\frac{\sigma}{b\sqrt{s}}, \label{eqn: ABC Kappa Def}\\
\mu = l+\half. 
\label{eqn: ABC Mu Def}
\end{eqnarray}
Equation \bref{eqn:Whittaker Equation} has Whittaker $M$ and $W$ functions as a satisfactory pair of solutions \cite{slater1960confluent} meaning the general solution is
\begin{eqnarray}
\hat{Q}(r,s) = AM_{\kappa,\mu}(z) + BW_{\kappa,\mu}(z).
\label{eqn:Whittaker General Solution}
\end{eqnarray}
As the Laplace transform of boundary condition \bref{eqn: BC SP exterior} is evaluated at infinity, its application can be achieved by inspection of appropriate asymptotic series. Assuming $c>0$ in the Bromwich integral implies that $-\half\pi<\arg{z}=\arg{br\sqrt{s}}< 0$ along the integration path, so the following equations are valid \cite{slater1960confluent} for $z\to\infty$:
\begin{dmath}[compact]
M_{\kappa,\mu}(z) 
\sim 
\frac{\Gamma(1+2\mu)}{\Gamma(\half+\mu-\kappa)}
z^{-\kappa}e^{\half z} \phantom{}_2F_0\brac{\half + \mu + \kappa,\half-\mu+\kappa,\frac{1}{z}} 
+
\frac{\Gamma(1+2\mu)}{\Gamma(\half+\mu+\kappa)} x^\kappa e^{-\half z}e^{\pi i(\kappa-\mu-\half)}  \phantom{}_2F_0\brac{\half + \mu - \kappa,\half-\mu-\kappa,-\frac{1}{z}}
\label{eqn: Whittaker M assym}
\end{dmath}
and
\begin{dmath}[compact]
W_{\kappa,\mu}(z) \sim z^{\kappa}e^{-\half z} \phantom{}_2F_0\brac{\half + \mu - \kappa,\half-\mu-\kappa,-\frac{1}{z}},
\label{eqn: Whittaker W assym}
\end{dmath}
where
\begin{eqnarray}
 \phantom{}_2F_0\brac{a_1,a_2,z} = \sum_{n=0}^\infty \frac{(a_1)_n(a_2)_n}{n!}z^n.
\end{eqnarray}
The Pochhammer notation, $(a)_n \equiv a(a+1)(a+2)\hdots(a+n-1)$ with $(a)_0=1$ has been used. The dominant terms in equations \bref{eqn: Whittaker M assym} and \bref{eqn: Whittaker W assym} are the exponential functions $e^{\half z}$ and $e^{-\half z}$ respectively. As $\Re(z)>0$ along the integration path then $z\to\infty$ as $r\to\infty$, so we must enforce $A=0$, in \bref{eqn:Whittaker General Solution}, in order for the boundary condition to be satisfied. So
\begin{eqnarray}
\hat{Q}(r,s) = B W_{\kappa,\mu}(br\sqrt{s}).
\end{eqnarray}
Division of the above by its derivative and rearranging produces
\begin{eqnarray}
\hat{Q}(r,s) = \frac{1}{b\sqrt{s}}\brac{\frac{W_{\kappa,\mu}(br\sqrt{s})}{\pderiv{W_{\kappa,\mu}(br\sqrt{s})}{r}}}\pderiv{\hat{Q}(r,s)}{r}.
\end{eqnarray}
Use of the convolution theorem \cite{transMethDuffy} and evaluating the result on $r=R$ yields the absorbing boundary condition,
\begin{eqnarray}
{Q}(r,t) = \int_0^t G_{\kappa,\mu}(R,\tau) \pderiv{{Q(R,t-\tau)}}{r} \, d\tau,
\label{eqn: Gen ABC}
\end{eqnarray}
where
\begin{eqnarray}
\hat{G}_{\kappa,\mu}(R,s) = \left.\frac{1}{b\sqrt{s}}\brac{\frac{W_{\kappa,\mu}(br\sqrt{s})}{\pderiv{W_{\kappa,\mu}(br\sqrt{s})}{r}}}\right|_{r=R}.
\label{eqn: general ABC kernel}
\end{eqnarray}
Once the inverse Laplace transform has been calculated to yield $G_{\kappa,\mu}(R,t)$ from $\hat{G}_{\kappa,\mu}(R,s)$, equation \bref{eqn: Gen ABC} can be discretised for use with the Crank-Nicholson scheme described in section \ref{subsec:Numerical Procedure}. We also note that \bref{eqn: Gen ABC} is non-local, meaning it depends on wavefunction information from previous times, which will be seen to have consequences for its numerical implementation described later. To proceed to find the inverse Laplace transform an implementation of a non-linear least squares method is used.

\subsection{\label{subsec:Bootstrap Method of Non-Linear Least Squares}Laplace Inversion of the Kernels}

Previously \cite{PhysRevC.87.014330} we relied on deriving a partial fractions representation, for which there is a known inversion. Surveying the literature \cite{AbraMathFunc,NIST:DLMF,slater1960confluent} it can be seen that the same technique cannot be applied to the kernel \bref{eqn: general ABC kernel}. Finding an exact inversion, then, appears unlikely. However, if we were to have an accurate approximate of the kernel given as a sum of some partial fractions, then an analytic inversion of the approximation could be performed. This can be achieved via a method of non-linear least squares \cite{Alpert00rapidevaluation}, where the mean square error,
\begin{eqnarray}
\int_a^b \left| \frac{P_d(z)}{Q_d(z)} - f(z) \right|^2 \, dz,
\end{eqnarray}
between a rational function, $\frac{P_d(z)}{Q_d(z)}$, and a kernel function, $f(z)$, is minimised. $P_d(z)$ and $Q_d(z)$ are polynomials of degree $d-1$ and $d$ respectively and $a$ and $b$ are two purely imaginary numbers. The rational function can be expressed as the sum-of-poles,
\begin{eqnarray}
\frac{P_d(z)}{Q_d(z)} 
= 
\sum_{k=1}^d \frac{w_k}{z-z_k}.
\end{eqnarray}
The Laplace inversion of the above is known to be the sum of exponentials \cite{AbraMathFunc},
\begin{eqnarray}
\mathcal{L}^{-1}\left\{\frac{P_d(z)}{Q_d(z)}\right\}
=
\sum_{k=1}^d w_ke^{z_k\tau}.
\end{eqnarray}
Finding an inverse Laplace transform is then reduced to calculating the values of the kernel function. However, for a Schroedinger equation's ABC kernel it was shown \cite{Jiang2001} that the method described in \cite{Alpert00rapidevaluation} could not be applied directly. This is due to the fact that more poles than can be calculated accurately with a numerical implementation of \cite{Alpert00rapidevaluation} are required to approximate it. As the kernel studied here also results from a Schroedinger equation then we expect the same to occur and so the modified bootstrap procedure given in \cite{Jiang2001,XuJiangBootstrap2013} is used. We proceed by describing the non-linear least squares method, before explaining how it is embedded into a bootstrap procedure.

\subsubsection{\label{subsubsec:Method of Non-Linear Least Squares}Method of Non-Linear Least Squares}

We require a method for finding polynomials $P_d(z)$ and $Q_d(z)$, where $d=\deg{P_d(z)}+1=\deg{Q_d(z)}$, such that
\begin{eqnarray}
\int_a^b \left| \frac{P_d(z)}{Q_d(z)} - f(z) \right|^2 \, dz
\end{eqnarray}
is minimised. As in the solution to the stationary Hartree-Fock equation, self consistency is used to linearize the problem and produce
\begin{eqnarray}
\int_a^b\left| \frac{P^{(i+1)}_d(z) - f(z)Q^{(i+1)}_d(z)}{Q^{(i)}_d(z)} \right|^2 \, dz,
\label{eqn: Linearized Least Squares}
\end{eqnarray}
where $i\ge1$ and
\begin{eqnarray}
&& P^{(i+1)}_d(z) = \sum_{j=0}^{d-1}  p_j z^j,\\
&&Q^{(i+1)}_d(z) = z^d + \sum_{j=0}^{d-1} q_j z^j.
\end{eqnarray}
The scheme in equation \bref{eqn: Linearized Least Squares} requires an initial guess, $Q^{(1)}_d(z)$, which we will describe later. It is hoped as we iterate through $i$ finding a minimum of \bref{eqn: Linearized Least Squares}, then the differences between the values of $\frac{P^{(i+1)}_d(z)}{Q^{(i+1)}_d(z)}$ and $f(z)$ become small. 

To minimise equation \bref{eqn: Linearized Least Squares}, $2d$ freedoms are introduced for the coefficients of $P_d^{(i+1)}(z)$ and $Q_d^{(i+1)}(z)$ which can be shown to produce the equations
\begin{eqnarray}
\int_a^b \bar{z}^n \frac{P^{(i+1)}_d(z)-f(z)Q^{(i+1)}_d(z)}{|Q^{(i)}_d(z)|^2} \, dz = 0 ,
\label{eqn: LS ortho 1}\\
\int_a^b \bar{z}^n \bar{f}(z) \frac{P^{(i+1)}_d(z)-f(z)Q^{(i+1)}_d(z)}{|Q^{(i)}_d(z)|^2} \, dz = 0,
\label{eqn: LS ortho 2}
\end{eqnarray}
where $n=1,\hdots,2d$, as sufficient conditions for a minimum. Defining the weighted inner product
\begin{eqnarray}
\langle f \mid g \,\rangle
=
\int_a^b \frac{\bar f(z)g(z)}{|Q^{(i)}_d(z)|^2} \, dz
\end{eqnarray}
and the basis
\begin{eqnarray}
h_n(z) = 
\begin{cases}
z^{\frac{n-1}{2}} f(z), & n = 1,3,\hdots,2d+1 \\
z^{\frac{n}{2}-1}, & n = 2,4,\hdots , 2d
\label{eqn:Least Squares Basis}
\end{cases},
\end{eqnarray}
allows equation \bref{eqn: LS ortho 1} and \bref{eqn: LS ortho 2} to be written simply as
\begin{eqnarray}
\langle h_n \mid -P+fQ \, \rangle = 0,
\end{eqnarray}
for $n=1,\hdots,2d$. We see from the above that the numerator of \bref{eqn: Linearized Least Squares} is orthogonal to the first $2d$ elements of the basis \bref{eqn:Least Squares Basis}. By inspection we can see that the numerator is also a linear combination of the entire basis. So orthogonalising the $2d+1$ functions in \bref{eqn:Least Squares Basis} will result in $-P(z)+f(z)Q(z)$ being the member of the resultant orthogonal basis that is in the span of $h_{2d+1}(z)$.

The restatement of this problem means we can apply the Gram-Schmidt process. This takes any set of linearly independent functions and produces a set of orthogonal functions, $g_n(z)$. The first two orthogonal functions given by the Gram-Schmidt process are
\begin{eqnarray}
&&g_1(z) = h_1(z), \\
&&g_2(z) = h_2(z) - \frac{\braket{g_1}{h_2}}{\braket{g_1}{g_1}}g_1(z).
\end{eqnarray}
Now instead of proceeding by orthogonalizing the set $\{ h_1(z),h_2(z),\hdots,h_{2d+1}(z) \}$, we take advantage of the Gram-Schmidt process being applicable to any set of linearly dependent functions. The basis holds the property $h_n(z) = zh_{n-2}(z)$ allowing us to continue by orthogonalizing the set $\{ h_1(z),h_2(z),zg_1(z),zg_2(z),\hdots,zg_{2d-1}(z) \}$, meaning for $n>2$
\begin{eqnarray}
g_n(z) = zg_{n-2}(z) - \sum_{j=1}^{n-1} \frac{\braket{g_j}{zg_{n-2}}}{\braket{g_j}{g_j}}g_j(z).
\label{eqn:Gram-Schmitt n>3 temp}
\end{eqnarray}
This simplifies the orthogonalization and results in a recursion in terms of just $g_n(z)$ with two initial values. It can be shown that $\{ h_1(z),h_2(z),zg_{1}(z),\hdots,zg_{n-2}(z) \}$ spans the same space as $\{ h_1(z),h_2(z),\hdots,h_n(z) \}$ for all $n\ge3$ via induction \cite{Pardi2013}. 

A final simplification is made by considering the value of the quantity $\braket{g_j}{zg_{n-2}}$ along the integration path, where
\begin{dmath}[compact]
\braket{g_j}{zg_{n-2}}
= 
-\braket{zg_j}{g_{n-2}}.
\end{dmath}
By insertion of equation \bref{eqn:Gram-Schmitt n>3 temp}, the quantity on the right hand side of the above can then be shown to be the following
\begin{dmath}[compact]
\braket{zg_j}{g_{n-2}} = \overline{\brac{\braket{g_{n-2}}{g_{j+2}} + \sum_{k=1}^{j+1} \frac{\braket{g_{k}}{zg_{j}}}{\braket{g_k}{g_k}}
\braket{g_{n-2}}{g_{k}} }}
\end{dmath}
By the orthogonality of the functions $g_n(z)$ we see, from the equations above, that
\begin{eqnarray}
\braket{g_j}{zg_{n-2}} = 0,
\end{eqnarray}
for $j<n-4$. The index of the sum in equation \bref{eqn:Gram-Schmitt n>3 temp} now runs from $n-5$ to $n-1$.  Re-expressing the sum so that the index runs from $1$ to $4$ means the Gram-Schmidt process can be summarised as
\begin{dmath}[compact]
g_n(z) = 
\begin{cases}
f(z), & n=1\\
1 - c_{21}g_1(z), & n=2 \\
zg_{n-2}(z) - \sum_{j=1}^{\min(n-1,4)} c_{nj}g_{n-j}(z), & n\ge3
\label{eqn:Gram-Schmitt Recur}
\end{cases},
\end{dmath}
where
\begin{eqnarray}
c_{nj} = \frac{\braket{g_{n-j}}{zg_{n-2}}}{\braket{g_{n-j}}{g_{n-j}}}.
\label{eqn:cnj Def}
\end{eqnarray}
We see $g_{2d+1}(z)=-P_d^{(i+1)}(z)+f(z)Q_d^{(i+1)}(z)$, as this is the only member in the span of $h_{2d+1}(z)$. Comparing with equation \bref{eqn: Linearized Least Squares}, 
\begin{eqnarray}
\braket{g_{2d+1}}{g_{2d+1}},
\end{eqnarray}
is seen to be the mean square error. 

The recursion \bref{eqn:Gram-Schmitt Recur} is used to find the set of values $c_{nj}$ and the mean square error, then by letting $g^{P,Q}_n(z) = g_n^P(z) + f(z)g_n^Q(z)$ and setting it equal to \bref{eqn:Gram-Schmitt Recur} we can find $P^{(i+1)}_d(z)$ and $Q^{(i+1)}_d(z)$ by considering
\begin{eqnarray}
g_n^{P,Q}(z) = zg_{n-2}^{P,Q}(z) - \sum_{j=1}^{\min(n-1,4)} c_{nj}g_{n-j}^{P,Q}(z),
\label{eqn: P or Q Recur}
\end{eqnarray}
where
\begin{eqnarray*}
&g_1^P(z) = 0 \,,\quad g_2^P(z) = -1 \, ,\quad g_{2d+1}^P(z) = P^{(i+1)}_d(z),& \\
&g_1^Q(z) = 1 \,,\quad g_2^Q(z) = -c_{21} \, ,\quad g_{2d+1}^Q(z) = Q^{(i+1)}_d(z).&
\end{eqnarray*}
To find the pole weights, the derivative of $Q_d^{(i+1)}(z)$ is also required. Differentiation of \bref{eqn: P or Q Recur} provides this via the formula:
\begin{dmath}[compact]
(g_n^{Q})^{\prime}(z) = g_{n-2}^{Q}(z) + z(g_{n-2}^{Q})^{\prime}(z) - \sum_{j=1}^{\min(n-1,4)} c_{nj}(g_{n-j}^{Q})^{\prime}(z),\nonumber\\
\end{dmath}
where
\begin{eqnarray}
(g_1^{Q})^{\prime}(z) = 0\,,\quad (g_2^{Q})^{\prime}(z) = 0, \\ (g_{2d+1}^{Q})^{\prime}(z) = \deriv{Q_d^{(i+1)}(z)}{z}.
\end{eqnarray}
The above formulae give us the ability to calculate $P_d^{(i+1)}(z)$, $Q_d^{(i+1)}(z)$ and $\deriv{Q_d^{(i+1)}(z)}{z}$ at any point between $a$ and $b$ along the imaginary axis. This provides a way to calculate the $Q_d^{(i+1)}(z)$ for the next iteration and also to expand $\frac{P_d^{(i+1)}(z)}{Q_d^{(i+1)}(z)}$ as partial fractions,
\begin{eqnarray}
\frac{P_d^{(i+1)}(z)}{Q_d^{(i+1)}(z)} = \sum_{k=1}^d \frac{w_k}{z-z_k}.
\end{eqnarray}
Muller's method \cite{Muller1956} is used to provide the poles, $z_k$, which are the roots of $Q_d^{(i+1)}(z)$, while the residue theorem \cite{mathClsQuaPhy}  yields the weights,
\begin{eqnarray}
w_k = \frac{P_d^{(i+1)}(z_k)}{\deriv{Q_d^{(i+1)}(z_k)}{z}}.
\label{eqn:poleStrengths}
\end{eqnarray}
In the implementation of the least squares procedure the integral in equation \bref{eqn:cnj Def} is discretised with the extended trapezium rule \cite{press2007numerical}, using 41 points. We also choose to iterate equation \bref{eqn: Linearized Least Squares} through $i=1,i_{max}$ in all calculations, which is found to work well for $i_{max}=5$.

The least squares method is reliant on access to a reasonable initial guess of the denominator $Q^{(0)}(z)$. Reference \cite{Alpert00rapidevaluation} used a continued fraction representation to gain a denominator. However, for equation \bref{eqn: general ABC kernel} this isn't available due to the square root in its argument. An initial guess is found by realising that an approximation with $d$ poles has to be calculated before we know the mean squared error. However, looping through $d=1,2,\hdots$ until the error is reduced sufficiently, gives an automatic way to produce an approximation with a given error. This can also be used to generate an initial guess from the previous step using \cite{Jiang2001}
\begin{eqnarray}
Q^{(0)}_d(z) = 
\begin{cases}
z - \brac{\frac{b+a}{2} - i\frac{b-a}{5}}, & d=1\\
(z-2z_{d-1})Q^{(0)}_{d-1}(z), & d\ge2
\label{eqn: LS Inital Guess}
\end{cases}.
\end{eqnarray}
In the above $z_{d-1}$ is the zero furthest from the imaginary axis. Calculation of the above specifies the initial guess for $d=1$, after which the results from the approximation with $d-1$ poles is used to give the guess for an approximation with $d$ poles. Algorithm \ref{alg:Non-linear Least Squares Method} summarises the method so far.
\begin{algorithm}
\caption{Non-linear Least Squares Method}
\label{alg:Non-linear Least Squares Method}
\begin{algorithmic}
\State{$d=1$.}
\State{Set $Q^{(1)}_1(z) = z - \brac{\frac{b+a}{2} - i\frac{b-a}{5}}$.}
\While{$\braket{g_{2d+1}}{g_{2d+1}} < \epsilon\int_a^b |f(z)|^2\, dz$}
\For{$i=1,i_{max}$}
\State{Calculate the coefficients $c_{nj}$ and $g_{2d+1}(z)$ using \bref{eqn:Gram-Schmitt Recur}.}
\State{Calculate the poles by applying Muller's method to $Q_d^{(i)}(z)$, which is calculated using \bref{eqn: P or Q Recur}.}
\State{Use the poles to calculate $Q^{(i+1)}_d(z)$.}
\EndFor
\State{Calculate $Q^{(1)}_{d+1}(z)$ from \bref{eqn: LS Inital Guess}.}
\State{$d=d+1$.}
\State{Calculate the mean square error, $\braket{g_{2d+1}}{g_{2d+1}}$.}
\EndWhile
\State{Calculate each pole's weight using equation \bref{eqn:poleStrengths}.}
\State{Return the $d$ poles and their corresponding weights.}
\end{algorithmic}
\end{algorithm}

\subsubsection{Bootstrap method of non-linear least squares}

Now the ability to find a pole approximation on an interval has been gained, we look at how this can be embedded in the bootstrap procedure that allows for an accurate approximation to the kernel in equation \bref{eqn: general ABC kernel}.

The modification described in \cite{Jiang2001} is to split up the interval on the imaginary axis into sub-intervals on which the function is smooth enough to be approximated well by the least squares method. There are three considerations that must be made for this to be successful.

First, we require some way of joining the approximations made on each sub-interval. The values of an approximation made on a particular sub-interval are non-zero outside of it and so simply adding the results means each approximation will interfere with one another. This can be solved by specifying some order to make the approximations. Then, by approximating the kernel on the first sub-interval, we continue by making approximations of the kernel with the previous results subtracted on the subsequent sub-intervals. In this way the current approximation takes account of the previous ones and adding the resultants will approximate the kernel.

Secondly, imagining the function on the complex plane then it may be the case that, in an interval, the kernel can be well approximated by poles which make a larger contribution elsewhere on the imaginary axis. To see this, consider an approximation where some poles are far away from the interval along the imaginary line on which the approximation was calculated. Generally these contributions to the current sub-interval are small, but can be large for other intervals which the poles are close to. So, approximations on the following sub-intervals will also have to describe these poles, not just the kernel. We therefore want to ensure poles located far from the sub-interval they were calculated on are excluded. 

It may also be the case that the size of a sub-interval is much smaller than the absolute value of the real part of a pole. This time we would not expect the least square algorithm to calculate this pole accurately because its contribution is over a larger interval than what we are analysing. So poles holding this property are excluded to.

It seems sensible that only poles that are found near to the part of the imaginary axis they were calculated on should be included in the approximation. Therefore, a near pole is defined and we say that only the near poles should be included in the sum of poles approximation. Specifically a pole, $z'_k$, is defined a near pole on the interval $[-1,1]$ if
\begin{dmath}[compact]
12 \le \int_{-1}^1 \frac{1}{|x-z'_k|^2}\,dx 
\,=\,
\frac{1}{\Im{z'_k}}\sbrac{\arctan\brac{\frac{\Re{z'_k} + 1}{\Im{z'_k}} } - \arctan\brac{\frac{\Re{z'_k} - 1}{\Im{z'_k}}}}.
\label{eqn:Near Pole criterion}
\end{dmath}
The notation $z'_k$ used is to specify a pole, $z_k$, that has been scaled onto $[-1,1]$ using
\begin{eqnarray}
z'_k = \frac{ z_k - \half (a + b)} { \half (b - a) },
\end{eqnarray}

The criterion \bref{eqn:Near Pole criterion} describes an elliptical area surrounding the $[-1,1]$ interval \cite{Jiang2001}. The value on the left hand side describes how tightly the ellipse is to enclose the interval, where a larger value would describe a smaller area. A value of $12$ yields an ellipse that tightly surrounds the interval that has been found to be practical for our purposes.

Finally it may also be the case that a pole with a positive real part may also suffice to give a good approximation of the function on a sub-interval. Here we make an assumption that no poles with positive real part should be included, as this would violate the requirement that all poles should be to the left of the contour in the Bromwich integral \bref{eqn:inversionDef}.

\paragraph{Splitting the Imaginary Axis}

\begin{figure}[h!]
\centering
\resizebox{0.47\textwidth}{!}{\includegraphics{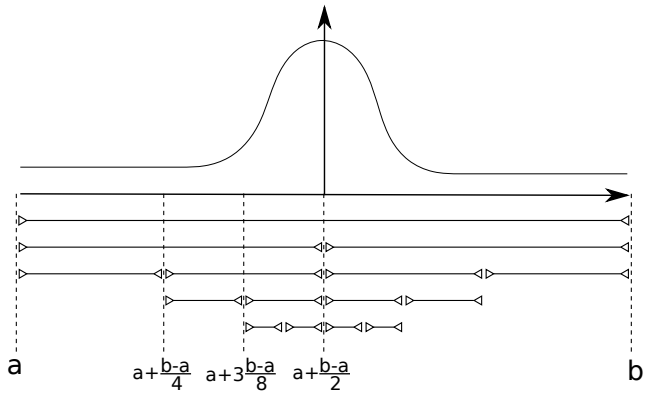}}
\caption{A diagram of how the binary splitting is expected to behave, with the smallest sub-intervals being created near the complicated behaviour of the function being approximated.}
\label{fig:splitInterval}
\end{figure}

Now a method to split the imaginary axis up into sub-intervals, on which the kernel can be approximated well by the least squares method, is required. This can be done recursively, by considering an interval and splitting it into two equally sized sub-intervals. On each of the sub-intervals a Chebyshev polynomial approximation is made and a criterion of whether or not this approximation is accurate has to be specified. If the criterion is satisfied, then no further splitting is done, but if it not, we split the sub-interval into two further sub-intervals and repeat the procedure. Figure \ref{fig:splitInterval} gives a simple illustration on how we would like a interval to be split up.

\subsubsection{The Splitting Criteria}

Deciding whether or not a kernel on a interval is smooth can be done on the assumption that if the function can be well approximated by a low order polynomial in that interval then the approximation will be successful. Therefore a Chebyshev polynomial approximation is calculated for the kernel on the interval and if the coefficients have got sufficiently small, implying convergence, we set the splitting criterion to false and the interval will not be split any further. A K-term Chebyshev approximation, $f_{\text{approx}}(z)$, of a function, $f(z)$, on $z\in[a,b]$ can be calculated using the formula \cite{press2007numerical},
\begin{eqnarray}
f_{\text{approx}}(z) = \sum_{k=0}^{K-1} \alpha_k T_k(z),
\label{eqn:Chebyshev Series}
\end{eqnarray}
The coefficients, $a_j$, are given by
\begin{eqnarray}
&&\alpha_0 = \frac{1}{K}\sum_{j=1}^{K} f(y_k), \\
&&\alpha_j = \frac{2}{K}\sum_{k=0}^{K-1} f(y_k)T_j(x_k).
\end{eqnarray}
The values of $x_k$, which are the roots of the Chebyshev polynomial, are given by
\begin{eqnarray}
x_k = \cos\brac{\frac{\pi(k+\half)}{n}}
\end{eqnarray}
and $y_k$ scales $x_k$ from $[-1,1]$ to $[a,b\,]$:
\begin{eqnarray}
y_k= \frac{(b+a) + (b-a)x_k}{2}.
\end{eqnarray}
If $f_{\text{approx}}(z)$ approximates $f(z)$ well, the coefficients of final terms in the series \bref{eqn:Chebyshev Series} should be relatively small. So, in practise we only require the coefficients and calculate
\begin{eqnarray}
S = \frac{ |\alpha_{K-1}| + |\alpha_{K-2}| }{\sum_{k=0}^{K-2}|\alpha_k|},
\end{eqnarray}
which can be thought of as a measure of convergence. A $\delta$ is defined so that if $S\le\delta$, the splitting criterion is set to false and if $S>\delta$ the splitting criterion is set to true. For all the results in this work the values $K=10$ and $\delta = 10^{-3}$ are used.

\paragraph{Binary Tree Description of an Interval}

Information on the sub-intervals is stored in a binary tree \cite{knuth1968art}. A binary tree is a collection of nodes which contain at least an association to a parent node and associations to left and right child nodes. These associations are called branches and give the tree its structure. It can be that a node's associations to both children are not specified, in which case we call it a leaf. There must be one and only one node without a parent, which we call the root. This defines a structure which has a single starting point, the root, and branches out to multiple endpoints, the leaves, like a tree. 

\begin{figure}[h!]
\centering
\resizebox{0.47\textwidth}{!}{\includegraphics{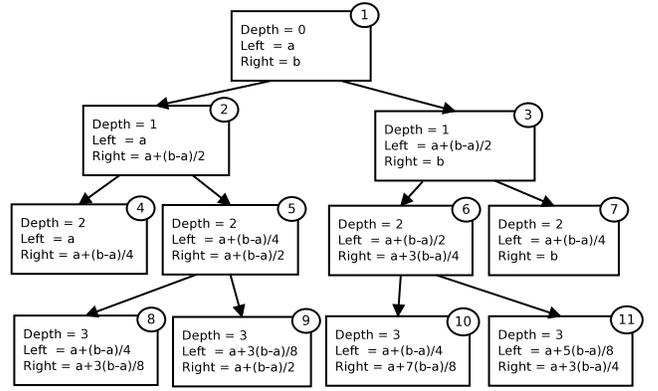}}
\caption{A diagram of the binary tree that describes the splitting, of the interval, in Figure \ref{fig:splitInterval}. Each box is a node with two arrows pointing away from it toward its children and a arrow from another node pointing towards it from its parent. Left and right specifies the two endpoints of the intervals.}
\label{fig:binaryTree}
\end{figure}

To make the binary tree useful for storing the splitting of our interval, we must append some additional information to each node. We choose to append the boundaries of each interval and what is called the node depth. The node depth is equal to the node depth of its parent plus one. The root's node depth is defined to be zero. For the interval split shown in figure \ref{fig:splitInterval} we would have a binary tree as shown in figure \ref{fig:binaryTree}.

\paragraph{Creating the Binary Tree}

Now we have shown how we may use a binary tree to describe the splitting of an interval we go on to describe how the binary tree is created. The procedure relies heavily on recursion and is described in algorithm \ref{alg:insert}.

\begin{algorithm}[!ht]
\caption{Insert(node)}
\label{alg:insert}
\begin{algorithmic}
	\If{node.depth $>$ maxdepth} maxdepth = node.depth
	\EndIf
	\If{(Splitting Criterion True)}
		\State{node.left.a = node.a}
		\State{node.left.b = $\half$(node.a+node.b)}
		\State{node.left.depth = node.depth + 1}
		\State{call insert(node.left)}
		\\
		\State{node.right.a = $\half$(node.a+node.b)}
		\State{node.right.b = node.b}
		\State{node.right.depth = node.depth + 1}
		\State{call insert(node.right)}
	\EndIf 
\end{algorithmic}
\end{algorithm}

Following the algorithm through we see that when a node is split in two, we move to its left child and check whether is needs to be split. If it does, two children are created and we move to the left again, if not then we move the parents right child and repeat the procedure. This process begins at the root of the tree.

\paragraph{The Bootstrap Method}

How the previous results are used to create an approximation to a kernel function is now specified. First create the binary tree, then begin at the left-most sub-interval at the maximum depth, and approximate the kernel and keep only the near poles. Then move rightwards through the rest of the nodes at that depth and approximate the kernel with all the previous near poles subtracted. Then move up to the next deepest and repeat the process, until the root is reached. Approximate the root, and keep all found poles. The order we would take for the tree shown in figure \ref{fig:binaryTree} would be 8, 9 ,10, 11, 4, 5, 6, 7, 2, 3, 1. To return the poles at a certain depth a modified in-order tree transversal \cite{knuth1968art}, as shown in algorithm \ref{alg:NodesAtDepth}, is used to produce a linked-list. The entire bootstrap procedure is summarised in algorithm \ref{alg:Bootstrap}.

\begin{algorithm}[!ht]
\caption{NodesAtDepth(node)}
\label{alg:NodesAtDepth}
\begin{algorithmic}

	\If{node.left exists}
		\State{NodesAtDepth(node.left)}
	\EndIf

	\If{$\text{node.depth} = d$}
		\State{Add node to end of linked list}
	\EndIf

	\If{node.right exists}
		\State{NodesAtDepth(node.right)}
	\EndIf

\end{algorithmic}
\end{algorithm}

\begin{algorithm}[!ht]
\caption{Bootstrap non-linear least squares}
\label{alg:Bootstrap}
\begin{algorithmic}
\State{Specify an interval $[a,b\,]$}
\State{Use algorithm \ref{alg:insert} to create a binary tree}
\For{$d=\text{maxdepth}:-1:1$}
\State{Use algorithm \ref{alg:NodesAtDepth} to return $N$ sub-intervals $[a_n,b_n]$} at depth $d$
\For{$n=1,N$}
\State{Use algorithm \ref{alg:Non-linear Least Squares Method} return the poles and corresponding weights} on $[a_n,b_n]$
\State{Discard poles and weights which don't meet criterion \bref{eqn:Near Pole criterion}}
\State{Add remaining poles and weights to list}
\EndFor
\EndFor
\State{Use algorithm \ref{alg:Non-linear Least Squares Method} to return poles and corresponding weights on root interval $[a,b]$ and add to list}
\State{Return list of weights and poles.}
\end{algorithmic}
\end{algorithm}

\subsection{\label{subsec:Boundary Kernel Calculation}Boundary Kernel Calculation}

How we calculate the proton kernel,
\begin{eqnarray}
\hat{G}_{\kappa,\mu}(R,s) = \left.\frac{1}{b\sqrt{s}}\brac{\frac{W_{\kappa,\mu}(br\sqrt{s})}{\pderiv{W_{\kappa,\mu}(br\sqrt{s})}{r}}}\right|_{r=R},
\label{eqn:Proton Kernel 2}
\end{eqnarray}
is now described. The kernels studied in \cite{Jiang2001} and \cite{Alpert00rapidevaluation} had continued fraction representations which provided an efficient and accurate means to calculate values over the entire complex plane. We choose a similar strategy and use the continued fraction \cite{NIST:DLMF},
\begin{eqnarray}
\frac{W_{\kappa,\mu}(z)}{\sqrt{z}W_{\kappa-\half,\mu-\half}(z)} =  
1 + \cfrac{\frac{v_1}{z}}{1+ \cfrac{\frac{v_2}{z}}{1+\hdots} },
\label{eqn:Whittaker Cont Frac}
\end{eqnarray}
where
\begin{eqnarray}
v_{2n+1} = \half + \mu - \kappa + n,\\
v_{2n}   = \half - \mu - \kappa + n.
\end{eqnarray}
The above converges for $|\arg(z)|<\half$ and $\mu+\half\pm(\kappa+1) \ne -1,-2,\hdots$. From section \ref{subsec:Absorbing Boundary Conditions} we know $-\half\pi<\arg{z}< 0$ and so the above equation is valid for our considerations. The use of the recurrence relation \cite{NIST:DLMF,slater1960confluent,Olver:2010:NHMF},
\begin{dmath}
W_{\kappa-\half,\mu-\half}(z)
=
\frac{1-2m-z}{(1-2m-2\kappa)\sqrt{z}}W_{\kappa,\mu}(z)
+
\frac{\sqrt{z}}{\kappa-\half+m}\deriv{W_{\kappa,\mu}(z)}{z} \,\,,
\end{dmath}
allows us to express \bref{eqn:Whittaker Cont Frac} in terms of a Whittaker function and its derivative as given in the kernel. The following continued fraction can then be written for the kernel:
\begin{dmath}[compact]
\hat{G}_{\kappa,\mu} (R,s)
= 
\frac{1}{b\sqrt{s}}
\cfrac{2z}{1-2\mu-z+\cfrac{2(\kappa+\mu)-1}{
1 + \cfrac{\frac{v_1}{z}}{1+ \cfrac{\frac{v_2}{z}}{1+\hdots} }}},
\label{eqn:kernelContFrac}
\end{dmath}
using $z$ as defined below equation \bref{eqn:LaplaceODE}. We calculate the above using Lentz's algorithm \cite{press2007numerical,Lentz:76,Thompson1986490}.

\begin{figure}[ht]
\resizebox{0.47\textwidth}{!}{\includegraphics{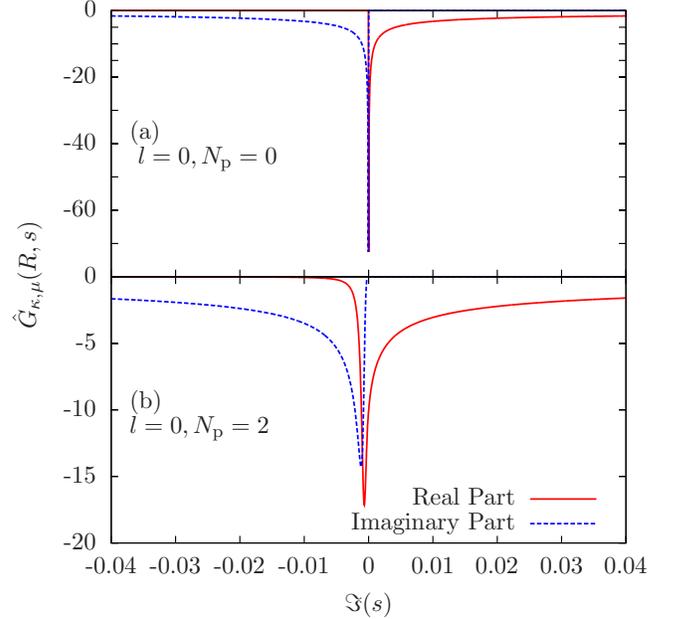}}
\caption{(Color online) A plot showing the values of the kernel \bref{eqn:Proton Kernel 2} using the values shown and $R=9.9$.}
\label{fig:Whittaker Kernel}
\end{figure}

In figure \ref{fig:Whittaker Kernel} the kernel is plotted for two different parameter sets: $l=0$, $N_{\text{p}}=0$ and $l=0$, $N_{\text{p}}=2$.
\\
Figure \ref{fig:Whittaker Kernel} shows that the complicated behaviour of the functions is centred nearby the origin and that the $N_{\text{p}}=0$ kernel appears less smooth than the one of $N_{\text{p}}=2$. This occurs when $N_{\text{p}}=0$ with small $l$ and will be shown to have consequences when the approximations are made for these kernels. Extending the plot's x-axis outwards would shows the function slowly decaying to zero.

\subsection{\label{subsec:Boundary Discretization}Boundary Discretization}

How we discretise the ABC with a proton kernel is now described. As $G_{\kappa,\mu}(R,\tau)$ and $\pderiv{{Q(R,t-\tau)}}{r}$ are both continuous we expect their numerical integration, by the trapezium rule, to be accurate. However, we have observed this is not the case when the trapezium rule is applied directly. This can be resolved by considering the case of $\sigma= 0$ where it is known the analytic form of the kernel contains a square root singularity \cite{springerlink:10.1134/S1054660X10050063}. The sum-of-exponentials should accurately describe this behaviour and hence not be integrated accurately by the trapezium rule. So, in analogy with \cite{PhysRevC.87.014330} integration by parts is used on equation \bref{eqn: Gen ABC}, before it is discretized, to produce
\begin{dmath*}[compact]
Q(R,t) + \pderiv{Q(R,t)}{r}H_{\kappa,\mu}(R,0)
=
-\int_0^t H_{\kappa,\mu}(R,\tau) \pderiv{}{\tau}\brac{\pderiv{Q(R,t-\tau)}{r}} \, d\tau.
\end{dmath*}
During the by-part manipulation $G_{\kappa,\mu}(R,\tau)$ is integrated to give
\begin{eqnarray}
H_{\kappa,\mu}(R,\tau) = \sum_{k=1}^d \frac{w_k}{s_k}e^{s_k\tau},
\label{eqn: int kernel}
\end{eqnarray}
which is also a sum-of-exponentials and can be easily evaluated.

\subsubsection{Time and Space Discretization}

A semi-discrete equation can be gained by evaluating $\tau$ at values on the temporal grid defined in section \ref{subsec:Numerical Procedure} for which $\tau=t_n$ and $t=t_N$. Use of the extended midpoint rule \cite{press2007numerical},
\begin{eqnarray}
\int_0^t f(\tau) \, d\tau = \Delta t \sum_{n=0}^{N-1} f\brac{t_{n+\half}} + \mathcal{O}(\Delta t^2),
\label{eqn: Trapezium time disc}
\end{eqnarray}
to evaluate the integral and the difference formulae;
\begin{eqnarray}
f(r,t_{n-\half}) = \frac{f(r,t_{n})+f(r,t_{n-1})}{2} + \mathcal{O}(\Delta t^2),\label{eqn: func time disc}\\
\pderiv{f(r,t_{n-\half}) }{t} = \frac{f(r,t_{n})-f(r,t_{n-1})}{\Delta t} + \mathcal{O}(\Delta t^2),
\label{eqn: deriv time disc}
\end{eqnarray}
for functions evaluated at a half time step, allows us to write the semi-discrete equation
\begin{dmath}
Q(R,t_N) + \pderiv{Q(R,t_N)}{r}H(R,0) = -\sum_{n=0}^{N-1} H(R,t_{n+\half}) \sbrac{\pderiv{Q(R,t_{N-n-1})}{r} -\pderiv{Q(R,t_{N-n})}{r} } + \mathcal{O}(\Delta t^2).
\end{dmath}
For the spatial discretization the absorbing boundary is applied at $R=r_{M-\half}$ between the penultimate and final spatial grid-points. The following difference formulae are used:
\begin{dmath}[compact]
f(r_{M-\half},t) = \frac{f(r_M,t)+f(r_{M-1},t)}{2} + \mathcal{O}(\Delta r^2), \label{eqn: func space disc}
\end{dmath}
\begin{dmath}[compact]
\pderiv{f(r_{M-\half},t) }{r} = \frac{f(r_{M},t)-f(r_{M-1},t)}{\Delta r} + \mathcal{O}(\Delta r^2), \label{eqn: deriv space disc}
\end{dmath}
at the points between the spatial grid, yielding the following fully-discretised ABC:
\begin{dmath}[compact]
\brac{1+B}Q(r_M,t_N) + \brac{1-B}Q(r_M,t_N) \\= -AH(r_{M-\half},t_\half)\Big(Q(r_M,t_{N-1}) - Q(r_{M-1},t_{N-1})\Big) 
- A\sum_{n=1}^{N-1} H(r_{M-\half},t_{n+\half})\Big(Q(r_M,t_{N-n-1}) - Q(r_{M-1},t_{N-n-1}) -Q(r_M,t_{N-n}) + Q(r_{M-1},t_{N-n})\Big)
+\mathcal{O}(\Delta r^2,\Delta t^2),
\label{eqn:fully disc proton ABC}
\end{dmath}
where
\begin{eqnarray}
&&A = \frac{2}{\Delta r},\\
&&B = A\Big(H(r_{M-\half},0) - H(r_{M-\half},t_\half)\Big).
\end{eqnarray}
Once the poles and weights have been calculated by using algorithm \ref{alg:Bootstrap}, they can be used with equation \bref{eqn: int kernel} to calculate the integral of the kernel for any required time. In general, we are required to recalculate algorithm \ref{alg:Bootstrap} for different values of $l$, $N_{\text{p}}$ and $R$. Replacement of the last equation of the matrix described in section \ref{subsec:Numerical Procedure} will then impose the boundary condition \bref{eqn: TDHF BC 2} on the calculation.

The fully discrete equation shows the consequence of the temporal non-locality of the ABC, noted at the end of section \ref{subsec:Absorbing Boundary Conditions}, as it contains a sum with upper bound $N-1$. This requires evaluating an increasing number of terms as the calculation progresses, which of course has implications on the computational cost. However, it has been noted that this can be remedied by a recursive evaluation of the absorbing boundary condition \cite{CPA:CPA20200}.

\section{\label{sec:Testing}Testing of the ABCs}

In this section the implementations of the bootstrap non-linear least squares and the absorbing boundary conditions are tested separately from the TDHF calculations. We start with various tests of the bootstrap implementation and then move on to show the results of applying the ABCs to some simple calculations of Schroedinger equations.

\subsection{\label{subsec:Testing of the Bootstrap Implementation}Testing of the Bootstrap Implementation}

\begin{figure}[!ht]
\centering
\resizebox{0.47\textwidth}{!}{\includegraphics{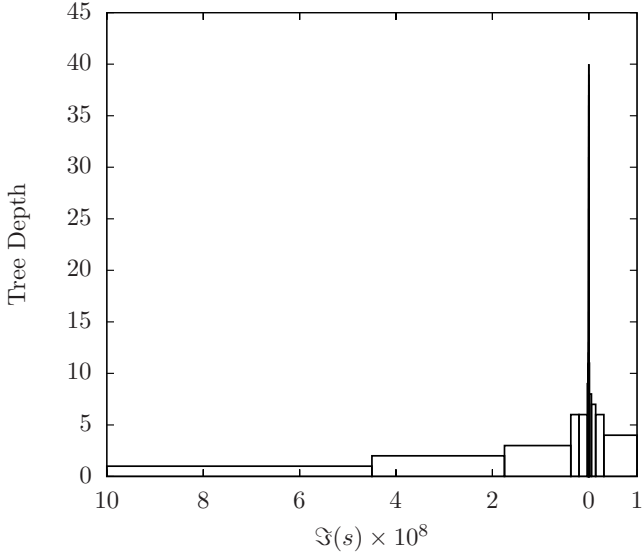}}
\caption{A box plot that shows the depths and sub-interval locations of the binary tree produced for a calculation using $l=0$ and $N_{\text{p}}=2$.}
\label{fig:intSplitting}
\end{figure}

The results of applying the least square approximation to the kernel \bref{eqn: general ABC kernel} are now shown. We have found that algorithm \ref{alg:Bootstrap} produces the smallest mean square error when used to make an approximation on an asymmetric interval. Therefore the values $a=-10^9i$ and $b=10^8i$ are chosen, so that the interval considered encloses the one used in \cite{Jiang2001} while being asymmetric. Firstly an example binary tree, produced by algorithm \ref{alg:insert}, is shown in figure \ref{fig:intSplitting} for $l=0$, $N_{\text{p}}=2$ and $R=9.9$.

By comparing figure \ref{fig:intSplitting} to figure \ref{fig:Whittaker Kernel} we can see that the width of the intervals become smallest around the complicated behaviour of the function, as wanted. It is noted that the tree depth is limited to $42$ in the implementation, as for $l=0$ and $N_p=0$ we want to prevent over-splitting, which may compromise the accuracy of the method. Figure \ref{fig:Whittaker Kernel} shows why this occurs, as the $N_{\text{p}}=0$ kernel is not as smooth as the $N_{\text{p}}\ne0$ kernel.

A selection of results is shown in table \ref{table:Bootstrap Errors} for values of $l$, $N_{\text{p}}$ and $R$ required by the Hartree Fock calculations.
\begin{table*}[!ht]
\centering
\begin{tabular*}{\textwidth}{@{\extracolsep{\fill} }  c  c  c  c  c  c}
 \hline\hline
 $R$ & $N_P$ & $l$ & No. of poles & Error on $[a,b]$ & Error on $[a,b]/(-10^{-4},10^{-4})$\\
 \hline
\multirow{9}{*}{ 9.9}
& 0&0&118&9.30$\times 10^{-2}$&2.56$\times 10^{-16}$\\
& 0&1&109&2.25$\times 10^{-13}$&1.36$\times 10^{-16}$\\
& 0&2&112&4.08$\times 10^{-14}$&6.53$\times 10^{-16}$\\
& 2&0&114&8.44$\times 10^{-17}$&7.92$\times 10^{-17}$\\
& 8&0&103&8.77$\times 10^{-17}$&8.66$\times 10^{-17}$\\
& 8&1&104&1.58$\times 10^{-16}$&1.59$\times 10^{-16}$\\
&20&0& 97&1.06$\times 10^{-16}$&1.05$\times 10^{-16}$\\
&20&1& 91&1.77$\times 10^{-16}$&1.77$\times 10^{-16}$\\
&20&2& 97&3.32$\times 10^{-16}$&3.33$\times 10^{-16}$\\
 \hline
\multirow{9}{*}{19.9}
& 0&0&117&9.30$\times 10^{-2}$&2.48$\times 10^{-16}$\\
& 0&1&117&7.73$\times 10^{-13}$&9.20$\times 10^{-17}$\\
& 0&2&112&1.46$\times 10^{-16}$&6.31$\times 10^{-17}$\\
& 2&0&108&1.15$\times 10^{-16}$&1.25$\times 10^{-16}$\\
& 8&0&104&7.77$\times 10^{-17}$&7.80$\times 10^{-17}$\\
& 8&1&108&7.03$\times 10^{-17}$&6.87$\times 10^{-17}$\\
&20&0&101&6.09$\times 10^{-17}$&6.06$\times 10^{-17}$\\
&20&1&100&7.37$\times 10^{-17}$&7.33$\times 10^{-17}$\\
&20&2& 92&6.73$\times 10^{-17}$&6.72$\times 10^{-17}$\\
 \hline
\multirow{9}{*}{29.9}
& 0&0&117&9.30$\times 10^{-2}$&2.50$\times 10^{-16}$\\
& 0&1&104&1.05$\times 10^{-11}$&1.50$\times 10^{-16}$\\
& 0&2&109&7.11$\times 10^{-15}$&1.95$\times 10^{-16}$\\
& 2&0&109&1.83$\times 10^{-16}$&1.96$\times 10^{-16}$\\
& 8&0&108&1.13$\times 10^{-16}$&6.91$\times 10^{-17}$\\
& 8&1&106&1.35$\times 10^{-16}$&1.29$\times 10^{-16}$\\
&20&0&102&2.54$\times 10^{-16}$&2.56$\times 10^{-16}$\\
&20&1& 98&1.21$\times 10^{-16}$&1.21$\times 10^{-16}$\\
&20&2&101&7.02$\times 10^{-17}$&6.94$\times 10^{-17}$\\
 \hline\hline
\end{tabular*}
\caption{Table showing the number of poles used to produce an approximation with the relative error specified. We show results for the values of $l$ and $N_{\text{p}}$ that are required by the Hartree-Fock calculations, for selected artificial boundaries. The first three approximations for each $R$ are used within calculations of the neutron single particle states and the remaining for the proton single particle states.}
\label{table:Bootstrap Errors}
\end{table*}

\begin{figure}[ht!]
\centering
\resizebox{0.47\textwidth}{!}{\includegraphics{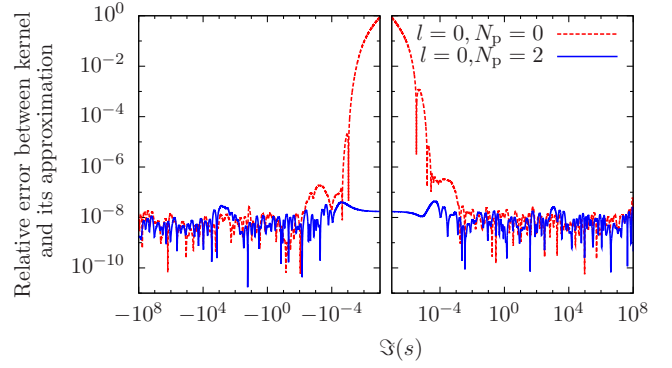}}
\caption{(Color online) A graph showing how the relative errors between two kernels and their approximations are distributed over the imaginary axis.}
\label{fig:errorOverImag}
\end{figure}

We see that for most cases the kernels are accurately approximated by the bootstrap method. Only for $N_{\text{p}} = 0$ does the mean square error become significantly larger and as $l$ increases the accuracy is recovered. This appears to be due to more complicated behaviour of the kernel around the origin, as shown in figure \ref{fig:Whittaker Kernel}, as the mean square errors with a small interval around the origin excluded are all similar. The larger error at the origin is confirmed by figure \ref{fig:errorOverImag} where examples are given to show how the relative error, between the approximation and the kernel, is distributed over the imaginary axis.

Both approximations have a similar magnitude of error away from the origin. However, in the approximation of $N_{\text{p}}$ we see that the error between the approximation and the kernel spikes. It will be shown later that the results presented are accurate enough for our needs.

\begin{figure}[!ht]
\centering
\resizebox{0.47\textwidth}{!}{\includegraphics{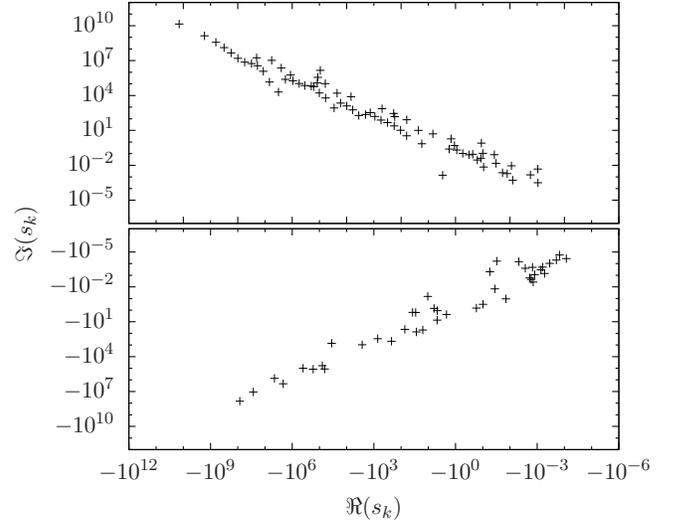}}
\caption{A graph of the complex plane, showing the pole locations found by the bootstrap method for the values $l=0$, $N_{\text{p}}=2$ and $R = 9.9$.}
\label{fig:polesLocation}
\end{figure}

Figure \ref{fig:polesLocation} shows the pole locations in the complex plane of the poles found by the bootstrap least square procedure . We denote the poles of the kernel \bref{eqn:Proton Kernel 2} by $s_k$.
\\
We see that the real and imaginary parts of the poles are similar in magnitude, which is a result of the near pole criterion.

\subsection{\label{subsec:Testing of the Absorbing Boundary Conditions}Testing of the Absorbing Boundary Conditions}

In this section the discretized ABC, equation \bref{eqn:fully disc proton ABC}, is tested for a simplified case of a Schroedinger equation with the same form within the interior as required by the exterior. Specifically the following will be solved:
\begin{dmath}[compact]
i\pderiv{Q_{l,N_{\text{p}}}(r,t)}{t} = \pderivtwo{Q_{l,N_{\text{p}}}(r,t)}{r} +\brac{ \frac{\eta N_{\text{p}}}{r} + \frac{l(l+1)}{r^2}}Q_{l,N_{\text{p}}}(r,t),
\end{dmath}
subject to the initial and boundary conditions
\begin{eqnarray}
&Q_{l,N_{\text{p}}}(r,0)= Are^{-(r-5)^2} ,& \label{eqn: Neutron test IC} \\
&Q_{l,N_{\text{p}}}(0,t)=0 \text{,\quad} \lim_{r\to\infty} Q_{l,N_{\text{p}}}(r,t) = 0. & \label{eqn: Neutron test BCs}
\end{eqnarray} 
In the above $A$ is chosen to normalise $Q_l(r,0)$. The values of $l$ and $N_{\text{p}}$ considered will be just those required by the TDHF calculations shown later.

\subsection{Radial Comparison of Wavefunction}

\begin{figure}[!htb]
\resizebox{0.475\textwidth}{!}{\includegraphics{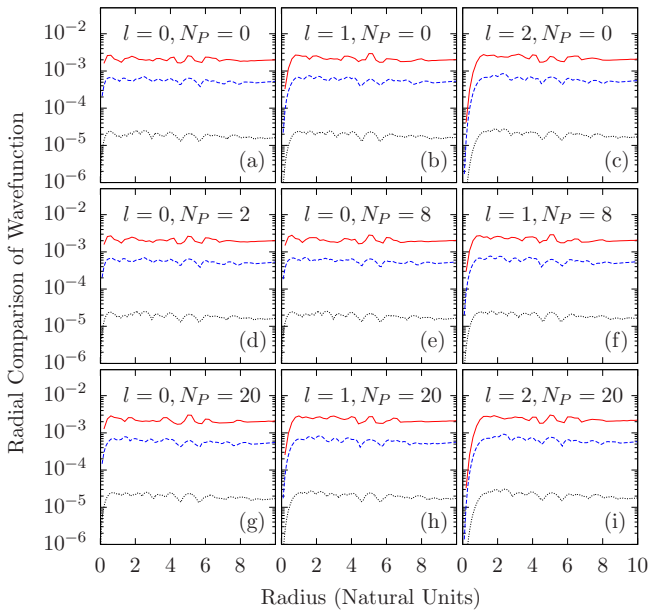}}
\caption{(Color online) The figures shows the maximum error of the radial component of the wavefunctions from times $0$ to $15$, for angular momenta and proton number shown, calculated with each technique. The value in equation \bref{eqn:Neutron plotMaxError} is plotted against the radius. The solid red lines show the result from using grid spacings $\Delta r=0.2$ and $\Delta t=0.2$, the dashed blue lines using $\Delta r=0.1$ and $\Delta t=0.1$ and the dotted black lines using $\Delta r=0.01$ and $\Delta t=0.01$.}
\label{fig:Proton plotMaxError}  
\end{figure}

Testing is begun by considering how the error from the absorbing boundaries affects the interior solution, by plotting the maximum absolute error that has occurred during the calculation. At each $r$
\begin{eqnarray}
\max_{t\in[0,50]}|Q^{(\text{Ref})}_{l,N_{\text{p}}}(r,t)-Q^{(\text{ABC})}_{l,N_{\text{p}}}(r,t)|
\label{eqn:Neutron plotMaxError}
\end{eqnarray}
is plotted, where $Q^{(\text{Ref})}_l(r,t)$ and $Q^{(\text{ABC})}_l(r,t)$ are the calculations with reflecting and absorbing boundaries respectively. Figure \ref{fig:Proton plotMaxError} shows the results for the various $l$ and $N_{\text{p}}$ values and three different grid spacings. The spacings $\Delta r=0.2$ and $\Delta t=0.2$ are chosen because it is the spacing we use in the Hartree-Fock calculations, the two other spacings are used to show the dependence of the error on the discretization. The reference solution is calculated on a grid with an outer boundary at 200 fm, which is far enough away to stop reflection occurring.
\\
We see that in all cases the error has remained small throughout the interior, for the $\Delta r=0.2$, $\Delta t=0.2$ case bounded by $10^{-2}$, for $\Delta r=0.1$, $\Delta t=0.1$ bounded by $10^{-3}$ and for $\Delta r=0.01$, $=\Delta t=0.01$ bounded by $10^{-5}$. The errors can be seen to be bounded similarly to those presented previously \cite{PhysRevC.87.014330}. There also appears to be no ill effects from the drop in accuracy, near the origin, of the $N_{\text{p}} =0$ approximations, with a similar magnitude of error being seen for all cases.  This is presumably due to the region of low accuracy being a rather small part of the whole region, and with sufficient unimportance to cause a serious problem.

\subsection{Temporal Comparison of Probability}

\begin{figure}[!ht]
\resizebox{0.475\textwidth}{!}{\includegraphics{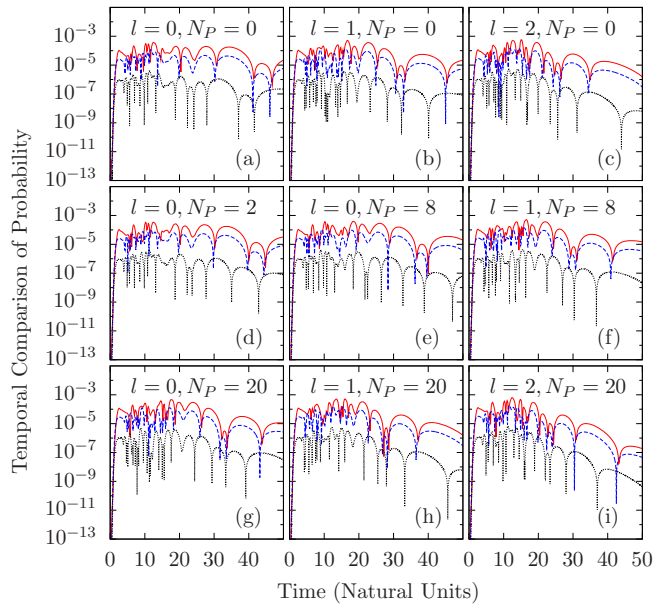}}
\caption{(Color online) These plots show how the error in the probability from the absorbing boundaries changes through time. Equation \bref{eqn: Neutron ProbTest} is calculated with reflecting and absorbing boundaries and the absolute value of their difference taken. The solid red lines show the result from using grid spacings $\Delta r=0.2$ and $\Delta t=0.2$, the dashed blue lines using $\Delta r=0.1$ and $\Delta t=0.1$ and the dotted black lines using $\Delta r=0.01$ and $\Delta t=0.01$.}
\label{fig:Proton normError}
\end{figure}

We now test how the error evolves through time. This is done by calculating the probability of finding a particle inside the interior region over time. Mathematically
\begin{eqnarray}
P(t) = \int_0^{10} |Q_{l,N_{\text{p}}}(r,t)|^2 \, dr
\label{eqn: Neutron ProbTest}
\end{eqnarray}
is calculated with reflecting and absorbing boundaries and the absolute value of the difference taken. Again the time interval of the calculation is $[0,50]$ and we choose the reflecting boundary to be at $r=200$. Figure \ref{fig:Proton normError} shows the results.
\\
We see that in time, also, the error remains bounded. From the plots it appears the bound on the error is proportional to the grid spacings. With the errors coming from the ABCs being small for all test cases, we proceed to use them with confidence.

\section{\label{sec:Results}Results from the TDHF}

\begin{figure}[!htb]
\centering
\resizebox{0.47\textwidth}{!}{\includegraphics{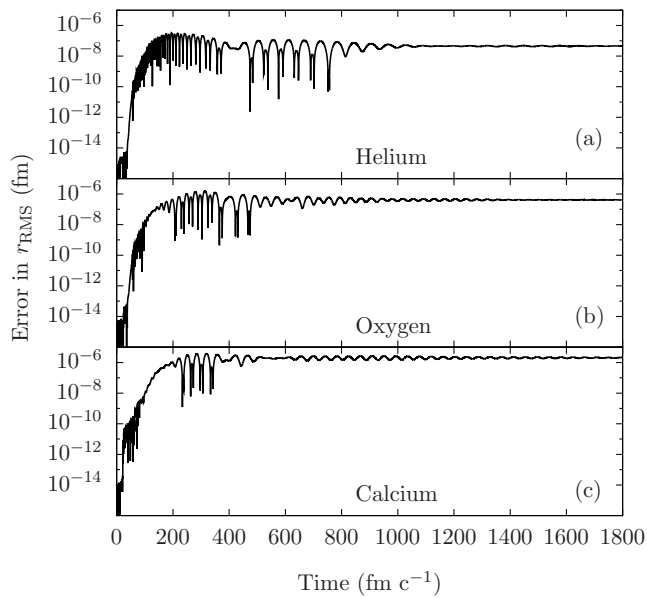}}
\caption{Plots showing the difference in the root mean square radii of reference calculations and a calculations on a grid with an outer boundary at 30 fm with ABCs applied.}
\label{fig:refRadsC_ABCerr}  
\end{figure}

In this section comparison will be made to reference solutions, calculated with reflecting boundaries on a grid with outer boundary at 700 fm to ensure reflection does not occur. Figure \ref{fig:refRadsC_ABCerr} shows the absolute error in the root mean square radius
\begin{eqnarray}
\brac{\int_0^8 4\pi r^4 \rho(r,t) \, dr}^\half,
\end{eqnarray}
between the reference solution and a calculation made with ABCs at 30 fm. Placement of the cutoff for the integration in the above is a parameter within TDHF calculations \cite{Stevenson2010}, but 8 fm appears to work well.

\begin{figure}[!htb]
\centering
\resizebox{0.45\textwidth}{!}{\includegraphics{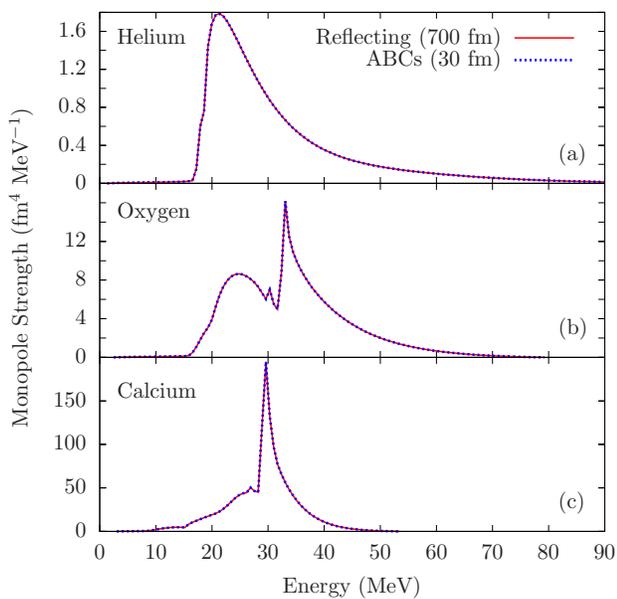}}
\caption{(Color online) Plots showing the strength functions for various nuclei. The solid red line shows the reference strength, whereas the blue dashes show the strength from a calculation on a grid with an outer boundary at 30 fm with ABCs applied.}
\label{fig:RefStrengthsC_ABCerr}  
\end{figure}

We see in each case the errors are consistently small and appear to be bounded by $10^{-5}$ fm. We therefore conclude that the non-linear portion of the potential is not large enough to disturb this type of calculation. However, these results will be of no use if the strength function is particularly sensitive to these errors and cannot be resolved properly. Therefore, a comparison of the reference strength functions with those calculated using ABCs is shown in figure \ref{fig:RefStrengthsC_ABCerr}.

The plots show this is not the case and the strength function calculated with ABCs is indistinguishable by eye to the reference. Therefore, strength functions from calculations using ABCs should be accurate enough to be successfully compared to experiment.

Finally the times taken to calculate the results and references are shown in figure \ref{fig:RefStrengthsC_ABCerr} are given in table \ref{table:Timing coulomb TDHF}, as we wish to see if there is any improvement in efficiency.
\begin{table}[!ht]
\centering
\begin{tabular}{l c c c c}
 \hline\hline
Nucleus & BNLS (s) & TDHF+ABCs (s) & Total (s) & Ref. Sol (s) \\
 \hline
  Helium&    1.99&    7.42&    9.41 & 135.16\\
  Oxygen&    2.79&   13.64&   16.43 & 267.38\\
  Calcium&    3.78&   24.12&   27.90 & 475.12\\
 \hline\hline
\end{tabular}
\caption{Table showing the time taken to calculate the various stages of the time-dependent code using ABCs. The values in the column labelled by BNLS are the times taken to calculate the bootstrap method of non-linear least squares for all kernels required, by the time-dependent Hartree-Fock calculation. The column labelled TDHF+ABCs shows the times taken to calculate the solution to the TDHF equations with ABCs applied at 30 fm. The column labeled total contains the sum of the times for the BNLS and TDHF+ABCs calculations. Finally, the values in the column labeled Ref. Sol are the time taken to calculate reference solution of the TDHF equations.}
\label{table:Timing coulomb TDHF}
\end{table}\\

The table shows us that the time for completion of each calculation has been drastically reduced. In this simplified case the absorbing boundary conditions approach has shown to be efficient and accurate.

\section{\label{sec:Conclusion}Conclusion and Outlook}

In this work we have presented an application of ABCs to Hartree-Fock calculations of spherical nuclei. ABCs with a centrifugal barrier and Coulomb potential in an exterior domain were considered. It was shown the ABCs required an inverse Laplace transform that was too complex for an analytical expression to be found. So, a bootstrap non-linear least squares method was implemented to produce an accurate sum-of-poles approximation to the kernel within the inverse Laplace transforms. The approximation was shown to be accurate and had an inverse known as a textbook result. Results of the ABC's application to TDHF calculations were similar to the Coulombless case considered previously \cite{PhysRevC.87.014330}, being accurate and efficient. 

On the physical side the outlook for this work is to include the full Skyrme interaction, allowing realistic calculations to be carried out \cite{Skyrme1958,doi:10.1080/14786435608238186}. It is expected that the ABCs would perform just as effectively for the full interaction, since the spitting between the interior and exterior regions is not affected. Secondly one would like to remove the restriction of spherical symmetry and extend the method to full three dimensional calculations \cite{PhysRevC.71.064328,brinemg,PhysRevC.73.054607}. This would allow other resonant modes to be studied and allow the calculation of non-spherical nuclei. Two possibilities under consideration to achieve this are via an expansion of the density in spherical harmonics, or an appropriate operator splitting method.

On the mathematical side we would like to offset some of the extra computational cost coming from physical improvements by increasing the efficiency of the implementation. A first improvement would be to change to a recursive evaluation of the ABCs, which is possible when using a sum-of-exponentials kernel \cite{CPA:CPA20200}. This would result in boundary conditions requiring just $\mathcal{O}(1)$ evaluations at each timestep. As well as this, an implementation of a temporal discretization scheme that is more suited to non-linear equations is desired. The literature already provides some methods applicable to the non-linear Schroedinger equation \cite{2005}, which offer the possibility to be generalised to the TDHF equations.

We conclude by remarking that the results presented highlight the ABC approach presented here as a valid method to handle the artificial boundary within TDHF calculations in the spatial basis. The current work also offers various avenues for improvement.

\bibliography{references}

\end{document}